\documentclass[10pt, conference, letterpaper]{IEEEtran}
\pdfoutput=1

\newcommand{\FULLPAPER}{} 

\ifCLASSOPTIONcompsoc
\usepackage[caption=false,font=normalsize,labelfon
t=sf,textfont=sf]{subfig}
\else
\usepackage[caption=false,font=footnotesize]{subfi
g}
\fi
\usepackage{lipsum,graphicx}
\usepackage{cite}
\usepackage{amsmath,amssymb,amsfonts}
\usepackage{algorithmic}
\usepackage{graphicx}
\usepackage{textcomp}
\usepackage{xcolor}
\usepackage{xspace}
\usepackage{enumerate}
\usepackage{subfig}
\usepackage{msc}
\setmsckeyword{} 

\usepackage{amsthm}
\usepackage{algorithm}
\usepackage{algorithmic}
\newtheorem{theorem}{Theorem}
\newtheorem{lemma}[theorem]{Lemma}
\newtheorem{corollary}[theorem]{Corollary}
\newtheorem{prop}[theorem]{Proposition}
\newtheorem{definition}[theorem]{Definition}
\newtheorem{problem}[theorem]{Problem}
\newtheorem{remark}[theorem]{Remark}
\newtheorem{protocol}[theorem]{Protocol}

\newcommand{\term}[1]{\emph{#1}\xspace}
\newcommand{\eg}{\emph{e.g.},\xspace}
\newcommand{\ie}{\emph{i.e.},\xspace}
\newcommand{\etal}{\emph{et al.}\xspace}
\newcommand{\cf}{\emph{cf.}\xspace}
\newcommand{\viz}{\emph{viz.}\xspace}
\newcommand{\A}{\ensuremath{\mathcal{A}}}
\newcommand{\EBC}[1]{\ensuremath{\mathrm{EBC}(#1)}\xspace}
\newcommand{\Pow}[1]{\ensuremath{\mathcal{P}\left(#1\right)}\xspace}
\newcommand{\Unif}[1]{\ensuremath{\mathrm{Unif}\left(#1\right)}\xspace}
\mathchardef\mhyphen="2D
\newcommand{\Xa}{\ensuremath{X^\mhyphen}\xspace}

\newcommand{\ForwardMessage}{\textsc{ForwardMessage}\xspace}
\newcommand{\ITSampler}{\textsc{InverseTransformSampler}\xspace}
\newcommand{\PFSampler}{\textsc{PickAndFlipSampler}\xspace}
\newcommand{\BackwardMessage}{\textsc{BackwardMessage}\xspace}
\newcommand{\PrivateEBC}{\textsc{PrivateEBC}\xspace}

\begin{document}

\title{Differentially-Private Two-Party \\ Egocentric Betweenness Centrality}

\author{\IEEEauthorblockN{Leyla Roohi\IEEEauthorrefmark{1},
Benjamin I. P. Rubinstein\IEEEauthorrefmark{2}, Vanessa Teague\IEEEauthorrefmark{3}
}
\IEEEauthorblockA{School of Computing and Information Systems,
University of Melbourne, Australia\\
Email: \IEEEauthorrefmark{1}lroohi@student.unimelb.edu.au,
\IEEEauthorrefmark{2}brubinstein@unimlb.edu.au,
\IEEEauthorrefmark{3}vjteague@unimelb.edu.au
}}
\thispagestyle{plain}    

\pagestyle{plain}
\maketitle
\begin{abstract}
We describe a novel protocol for computing the egocentric betweenness centrality of a node when relevant edge information is spread between two mutually distrusting parties such as two tele\-communications providers.  While each node belongs to one network or the other, its ego network might include edges unknown to its network provider.  We develop a protocol of differentially-private mechanisms to hide each network's internal edge structure from the other; and contribute a new two-stage stratified sampler for exponential improvement to time and space efficiency. 
Empirical results on several open graph data sets demonstrate practical relative error rates while delivering strong privacy guarantees, such as 16\% error on a Facebook data set. %
\end{abstract}


\section{Introduction}
Data sets such as social, communication, and transport networks are graph structured:  people are nodes and their interactions edges.  Such graph structures are valuable for understanding real-world properties.  However, revealing  the graph or its statistics can cause privacy disclosure even with anonymisation techniques~\cite{backstrom2007wherefore, narayanan2011link,narayanan2009anonymizing}.    
Furthermore, for many corporations, customer data is an asset they are reluctant to share. 
This motivates interest in joint computation over databases with limited exposure to sensitive information.  

Differential privacy (DP)~\cite{dwork2006calibrating} guarantees that a release output distribution does not change by more than a small multiplicative factor under  input data perturbation.
We consider \emph{edge DP} wherein perturbations correspond to edge flips: the existence of sensitive edges is not revealed by edge-DP release.    

We envisage two (or more) networks controlled by different corporations, such as telephone or email providers, or two different social networks.  The complete list of nodes (\ie people) is public knowledge, but the individual connections between them are not, so we consider edge DP in order to hide the connection between the nodes in each network. Each service provider knows the connections within its own network, plus the connections between one of its members and the outside (\eg when they contact someone in a different network), but not the internal connections in other networks.

We are the first to consider differentially-private computation of \emph{egocentric betweenness centrality} (EBC)~\cite{goh2003betweenness}. 
Informally, EBC measures the importance of a node as a link between different parts of the graph.  A node that forms a link between otherwise-isolated parts of the network has high betweenness centrality; a node that is simply an easily-bypassed member of an interconnected network has a low betweenness centrality.  This is a property of the whole communication graph: one service provider cannot compute it using its partial view of the graph alone.  

Betweenness centrality could be used  in targeted advertising or customer retention campaigns, as individuals with high EBC have the capacity to transfer information from one community to another.  EBC is equally important in understanding and combating the spread of misinformation or ``fake news'': individuals with high EBC can be educated to be more discerning about what they spread through the network, thereby mitigating spread of fake news.   
The difficulty of assessing misleading political content and obstructing its spread has become one of the most important research questions in online social network analysis, to which even the networks themselves are devoting significant research effort.\footnote{https://newsroom.fb.com/news/2018/04/new-elections-initiative/}

We enable a network provider to compute the egocentric betweenness centrality of a node, while requiring only differentially-private information about internal connections to be shared between networks. Our main contributions are:
\begin{enumerate}
    \item We introduce a privacy preserving method to compute the egocentric betweenness centrality of nodes in undirected graphs. In this work the network has local connections, while there are also inter-network connections. 
    \item We propose a two-stage sampling process that delivers a simple approach to implement and exponential  savings in time and space over na\"ive sampling from the exponential mechanism directly.
    \item We report on thorough experiments using a Facebook graph data set on 63,000 nodes. The experiments in Section~\ref{sec:experiments} show that the error is approximately 16\% of the true EBC for reasonable values of privacy level $\epsilon$.  Similar results hold for other networks from Enron and PGP email.
\end{enumerate}

First we survey the technical background and give a precise definition of EBC.  We then explain why a precise computation would expose individual links between networks.  Section~\ref{sec:proposed} describes our differentially-private mechanism for communicating enough information between networks to permit effective approximation of EBC while preserving strong privacy guarantees.  We then present empirical results testing the feasibility of our approach on samples of public data from Facebook, Enron and PGP.

\subsection{Related Work}

$k$-anonymity~\cite{sweeney2002k} represents a major, early attempt at preventing node and edge re-identification by graph transformation, but it has been proven to be insufficient~\cite{aggarwal2008general}.  

Differential privacy for graph processing was first introduced in \cite{hay2009accurate} and was followed up by further work~\cite{zhang2015private,day2016publishing,mulle2015privacy,shen2013mining}. 
Two main privacy models exist when publishing graph-based information under differential privacy; node~\cite{day2016publishing} and edge differential privacy~\cite{hay2009accurate}. 
Hay \etal \cite{hay2009accurate} introduced an  algorithm for publishing degree distributions under edge privacy, implicitly permitting private $K$-star counting as well. 
Projection-based techniques have been proposed to answer degree distribution queries under node differential privacy~\cite{kasiviswanathan2013analyzing,raskhodnikova2015efficient}.

Other statistics have been approximated under differential privacy such as frequent patterns of given sub graphs~\cite{zhang2015private,bhaskar2010discovering,karwa2011private}.  
Bhaskar \etal \cite{bhaskar2010discovering} used the exponential mechanism to publish the (approximately) most frequent patterns with high probability, and the Laplace mechanism to release the noisy frequency of maximising patterns. Karawa \etal \cite{karwa2011private} proposed a differentially-private algorithm to output answers to sub-graph counting queries for $K$-star, $K$-triangle sub graphs while using local sensitivity~\cite{nissim2007smooth} to overcome high global sensitivity in sub-graph counting queries. An approach to finding arbitrary frequent patterns was proposed by Shen \& Yu~\cite{shen2013mining}. They utilise the exponential mechanism and Markov chain Monte Carlo  
sampling to output frequent patterns on graph data sets.

Finding node clusters in a single graph under differential privacy was first proposed by \cite{mulle2015privacy} and followed by \cite{ nguyen2016detecting}. These techniques try to find the group of nodes sharing many links with other nodes in the same group but relatively few outside the group. They maintain the privacy of the output clusters under node or edge differential privacy.

Our work differs from previous studies in two key ways. 
First we focus on the problem of node influence, through the study of ego betweenness centrality. This particular task poses significant technical challenges, made efficient here by adopting two-stage stratified and accept-reject sampling. 
Second we consider a core graph processing task in a distributed two-party setting. While most existing work on graph mining under differential privacy can adopt a model of trusted computation, and there is some work on privacy for distributed systems~\cite{dwork2006our,chen2012towards},  these are based on distributed queries that are decomposed into sub-queries, each answered per database. Our setting requires untrusting parties to cooperate on computation without revealing one another's privacy-sensitive data.

\section{Preliminaries}  \label{sec:preliminaries}

\subsection{Egocentric Betweenness Centrality}

First proposed by Everett \& Borgatti \cite{everett2005ego} as an approximation to betweenness centrality~\cite{freeman1978centrality}, egocentric betweenness centrality (EBC) has gained recognition in its own right as a natural measure of a node's importance as a network bridge~\cite{marsden2002egocentric}.  The EBC of a node $a$ is the sum, for all pairs of neighbours of $a$ that aren't directly connected, of the fraction of 2-edge paths between them that pass through $a$.  


\begin{definition}\label{def2.2}
\term{Egocentric betweenness centrality (EBC)} of node $a$ in simple undirected graph $(V,E)$ is defined as 
\begin{eqnarray*}
\EBC{a} &=& \sum_{i,j\in N_a : A_{ij}=0, j>i}
\frac{1}{A^2_{ij}}\enspace,
\end{eqnarray*}
where $N_a=\{v\in V\mid \{v,a\}\in E\}$ denotes the neighbourhood or \term{ego network} of $a$, $A$ denotes the $(|N_a|+1)\times (|N_a|+1)$ adjacency matrix induced by $N_a\cup\{a\}$ with $A_{ij}=1$ if $\{i,j\}\in E$ and $0$ otherwise;  $A^2_{ij}$ denotes the
$ij$-th entry of the matrix square, guaranteed positive for all $i,j\in N_a$ since all such nodes are connected through $a$.
\end{definition}



\subsection{Differential Privacy on Graphs}
Differential privacy was proposed to quantify the indistinguishability of input databases when observing the output of data analysis \cite{dwork2006calibrating}.
With careful selection of which databases are to be indistinguishable---through the so-called neighbouring relation---the protective semantics of differential privacy may be controlled. 

As detailed further in Section~\ref{sec:problem}, our concern is maintaining the privacy of \emph{connections} in networks, \eg who calls whom in a telecommunications network. We therefore use \term{edge privacy} \cite{hay2009accurate} and so relate graphs that differ by edges. The adjacency matrix fully represents the edgeset of a graph of known nodes (the indices into the adjacency matrix). As such, we focus on databases as sequences of bits: elements of $\{0,1\}^n$. 

Formally, two databases $D,D'\in\{0,1\}^n$ and are termed \term{neighbouring} (denoted $D\sim D'$) if there exists exactly one $i\in [n]$ such that $D_i\neq D'_i$ and $D_j=D'_j$ for all $j\in [n]\backslash \{i\}$.
In other words, $\|D - D'\|_1=1$.

\begin{definition}
For $\epsilon>0$, a randomised algorithm on databases or \term{mechanism} $\mathcal{A}$ is said to preserve \term{$\epsilon$-differential privacy} if for any two neighbouring
databases $D, D'$, and for any measurable set $R\subseteq\mathrm{Range}(\mathcal{A})$,
\begin{IEEEeqnarray*}{rCl}
\Pr\left(\A(D) \in R\right) &\leq& \exp(\epsilon)\cdot \Pr\left(\A(D') \in R\right)\enspace.
\end{IEEEeqnarray*}
\end{definition}
\subsubsection{Generic Mechanisms for Privacy}
We leverage two well-known DP mechanisms in this paper: the Laplace mechanism~\cite{dwork2006calibrating} which applies additive noise to numeric vector-valued analyses, and the exponential mechanism~\cite{mcsherry2007mechanism} which privately optimises a real-valued objective function bivariate in the database and the decision variable which need not be numeric. Common to most generic mechanisms, and the Laplace and exponential in particular, is the concept of sensitivity-calibrated randomisation: the more sensitive a target function is to input perturbation, the more randomisation is required to attain a level of differential privacy. Both mechanisms leveraged here are calibrated via the same measure of sensitivity, defined next. 

\begin{definition}
The \term{$L_1$-global sensitivity} of any function \linebreak[4] $f: \{0,1\}^n\to\mathbb{R}^d$ for any $d\in\mathbb{N}$, is defined as
\begin{IEEEeqnarray*}{rCl}
\Delta f &\geq& \sup_{D,D'\in\{0,1\}^n, D\sim D'} \|f(D)-
f(D')\|_{1}\enspace.
\end{IEEEeqnarray*}
For functions of additional variables $f: \{0,1\}^n\times \Theta \to \mathbb{R}^d$ we extend this definition naturally as
\begin{IEEEeqnarray*}{rCl}
\Delta f &\geq& \sup_{\theta\in\Theta}\; \sup_{D,D'\in\{0,1\}^n, D\sim D'} \|f(D,\theta)-
f(D',\theta)\|_{1}\enspace.
\end{IEEEeqnarray*}
\end{definition}

We can now define the aforementioned generic mechanisms. 

\begin{lemma}\label{lem:laplace}
Consider any Euclidean vector-valued deterministic function $f: \{0,1\}^n \to \mathbb{R}^d$ for any $d\in\mathbb{N}$,  
and any scalar $\epsilon>0$. Given input $D\in\{0,1\}^n$,
the Laplace mechanism releases responses in $\mathbb{R}^d$ distributed as $f(D)+\mathbf{X}$ where
$\mathbf{X}$ is $d$ i.i.d. zero-mean Laplace\footnote{The zero-mean scalar Laplace with scale $\lambda>0$ has PDF $(2\lambda)^{-1} \exp(-|x|/\lambda)$.} r.v.'s with scale $\Delta f / \epsilon$. Then the Laplace 
mechanism preserves $\epsilon$-differential privacy.
\end{lemma}

\begin{lemma}\label{lem:exponential}
Consider any real-valued bivariate \term{quality function}
$q: \{0,1\}^n\times\Theta \to \mathbb{R}$, which assigns quality score $q(D,\theta)$ to candidate response $\theta\in\Theta$, on input database $D\in\{0,1\}^n$. The exponential mechanism
approximately maximises $q(D,\cdot)$ by releasing randomised response $\theta$ with likelihood proportional to $\exp( q(D,\theta) \cdot \epsilon / (2\cdot \Delta q))$. 
Then the exponential 
mechanism preserves 
$\epsilon$-differential
privacy. 
\end{lemma}

\subsubsection{Compositional Calculus}
In order to build up more complex privacy-preserving computations, it is necessary to be able to quantify the privacy loss of compositions. Fortunately, 
differential privacy satisfies sequential composition and transformation invariance \cite{dwork2006calibrating,kifer2010towards, mcsherry2009privacy} among other compositions.

\begin{lemma}[Sequential composition]\label{lem:composition}
For any sequence of \linebreak[4] randomised mechanisms $\mathcal{A}_1,\mathcal{A}_2,\ldots,\mathcal{A}_k$,
if each $\mathcal{A}_i$ preserves $\epsilon_i$-differential privacy then the compound response on a database $D$,  $(\A_1(D),\ldots,\A_k(D))$, preserves $\left(\sum_{i=1}^{k} \epsilon_i\right)$-differential privacy. 
\end{lemma}

\begin{lemma}[Transformation invariance]
For any mechanism $\A_1$ that is $\epsilon$-differentially private, and any (possibly randomised) mapping $\A_2$ with domain containing the co-domain of $\A_1$, the randomised mechanism $\mathcal{A}=\A_2\circ\A_1$ preserves $\epsilon$-differential privacy.
\end{lemma}

\section{Problem Statement}\label{sec:problem}

Consider a two-party setting of  a telecommunications network with two service providers $X, Y$: every customer is represented as a node $a$ that belongs to one and only one service provider; pairs of customers who \eg have called one another are represented as edges in a simple undirected graph on the disjoint union of nodes. Edges can either connect nodes within one party ($X$ or $Y$) in which case are unknown to the other party ($Y$ or $X$ respectively), or edges span both parties and are known to both. We consider all nodes to be known to both parties, as being addressable within a global addressing system (\eg a phone book).

Denote by $V_X, V_Y$ the nodes of $X,Y$ respectively, $E_X\subseteq V_X, E_Y\subseteq V_Y$ the edges (two-element sets) within parties $X, Y$ respectively, and $E_{XY}\subseteq V_X\cup V_Y$ the edges spanning $X,Y$ as sets with one element each from $V_X, V_Y$. The simple undirected graph on the entire network comprises node-set disjoint union $V_X\cup V_Y$ and edge-set disjoint union $E_X\cup E_Y\cup E_{XY}$. Note we will often equivalently represent edge sets as adjacency matrices (or flattened vectors) with elements in $\{0,1\}$. Table~\ref{tab:glossary} shows all of the symbols used in this paper.

We wish to enable one party (without loss of generality) $X$ to compute the ego betweenneess centrality (EBC) of one of its nodes $a\in V_X$, while maintaining \emph{edge privacy} between parties. 
Before detailing a protocol for accomplishing this task, we must be precise about a privacy model.
\begin{table}[t]
    \captionsetup{justification=centering, labelsep=newline}
    \renewcommand{\arraystretch}{1.3}
    \caption{Glossary of symbols used in this paper.}
    \label{tab:glossary}
    \centering
    \begin{tabular}{c|p{0.7\columnwidth}}
    \hline
    $X, Y$ & The two parties \eg competing service providers. \\
    $V_X, V_Y$ & The nodes per party. \\
    $E_X, E_Y$ & Edges entirely within each party. \\
    $E_{XY}$ & Edges spanning both parties. \\
    $a$ & The ego node (assumed WLOG to be in $X$). \\
    $N_a$ & The ego network of $a$. \\
    $\Xa$ & Party $X$'s nodes $V_X$ excluding $a$. \\
    $R^\star$ & The ego network contained in $X$. \\
    $T_{ij}$ & Counts of 2-paths spanning $X,Y$. \\
    $S_X, S_Y, S_{XY}$ & Partial EBC sums by endpoints. \\
    $R$ & A private, randomised approximation to $R^\star$. \\
    $Q_i$ & For $i\in\{0,\ldots,|\Xa|\}$, a partition of $\Pow{\Xa}$. \\
    $\epsilon$ & The differential-privacy budget. \\
    $\Delta$ & A global sensitivity bound. \\
    \hline
    \end{tabular}
    
\end{table}
\begin{problem}[Private Two-Party EBC] \label{prob:pebc}
Consider a simple undirected graph $(V_X\cup V_Y, E_X\cup E_Y\cup E_{XY})$ partitioned by parties $X,Y$ as above, and an arbitrary node $a\in V_X$. The problem of \term{private two-party egocentric betweenness centrality} is for the parties $X,Y$ to collaboratively approximate \EBC{a} under  assumptions that:
\begin{enumerate}[{A}1.]
    \item Both parties $X,Y$ know the entire node set $V_X\cup V_Y$; \label{ass:nodes}
    \item Each party knows every edge incident to nodes within their own network. That is, $X$ knows $E_X\cup E_{XY}$ while $Y$ knows $E_Y\cup E_{XY}$; and \label{ass:edges}
    \item The computed \EBC{a} needs to be available to $X$ but need not be shared with $Y$. \label{ass:ebc}
\end{enumerate}
Any solution must not reveal to $X,Y$ what is not already known except for $X$ discovering \EBC{a}  (Assumption~\ref{ass:ebc}). 
We seek solutions under an honest-but-curious adversarial model: while $X,Y$ will follow any agreed upon protocol prescribing computations to take and messages to send to one-another, without attempting to manipulate the other party; each party is curious about the other's edges and may apply arbitrary auxiliary computation and leverage data sources in attempting to discover the other's edges. Formally, what is revealed by $X$ ($Y$) to $Y$ (respectively $X$) must preserve $\epsilon$-differential privacy with respect to $E_X$ (respectively $E_Y$).
\end{problem}

\section{Warm-Up: A Non-Private Protocol}\label{sec:nonprivate}

We first consider how $X,Y$ might cooperate without preserving differential privacy.
In particular $X$ cannot itself count 2-paths that are
\begin{itemize}
    \item Contained entirely within $Y$; or
    \item Ending in both $X,Y$ with intermediate node in $Y$.
\end{itemize}
Any protocol must involve $Y$ in aggregating over such paths. But while the first case can be aggregated independently by $Y$, the second case requires $X$ to communicate its endpoint neighbours of $a$ to $Y$.
This significantly complicates the differentially-private solution developed in the next section.

\begin{figure}[t!]
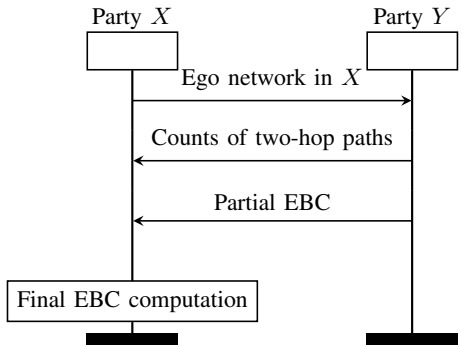

\centering
\vspace{-2em}
\drawframe{no}
\begin{msc}[small values, instance
distance=2.5cm,left environment distance=2.6cm,
right environment distance=1.9cm]{}
\declinst{usr}{Party $X$}{}
\declinst{m1}{Party $Y$}{}
\mess
{Ego network in $X$}{usr}{m1}
\nextlevel
\nextlevel
\mess{Counts of two-hop paths}{m1}{usr}
\nextlevel
\nextlevel
\mess{Partial EBC}{m1}{usr}
\nextlevel[2]
\action*{Final EBC computation}{usr}
\nextlevel
\end{msc}
\caption{EBC two-party computation protocol, comprising one forward and two backward messages.}
\label{fig:EBC}
\drawframe{yes}
\end{figure}

Recall that $N_a=\{v\in V \mid \{v,a\} \in E\}$ denotes the ego network of $a$ anywhere in the graph (notably not including $a$ since the graph has no self-loops).  Figure~\ref{fig:EBC} summarises the following protocol.

\begin{protocol} Proceeding in sequence:
\label{prot:NonPrivate}
\begin{enumerate}[i.]
   
\item {[Forward message]} $X$ sends to $Y$ the set $R^\star = N_a \cap V_X$ of neighbours of $a$ contained within $X$;
\label{step:nonPrivateProtForward}

\item {[Backward message]} $Y$ computes and sends to $X$, for each $i\in R^\star$ and for each $j\in N_a\cap V_Y$ (where $i,j$ are not directly connected), a count $T_{ij}$ of 2-paths with endpoints $i,j$ and intermediate point in
$N_a\cap V_Y$;  
\label{step:nonPrivateProtBack1}

\item {[Backward message]} $Y$ computes and sends to $X$, the EBC partial sum over endpoint nodes $i,j\in N_a\cap V_Y$ with intermediate nodes in $N_a\cup\{a\}$. That is, $S_Y=\sum_{i,j\in N_a\cap V_Y: A_{ij}=0, j>i} 1 / A_{ij}^2$; 
\label{step:nonPrivateProtBack2}

\item $X$ increments the received $T_{ij}$ by the number of 2-paths between $i,j$ with intermediate point in $R^\star \cup \{a \}$. It then sets $S_{XY}$ to the sum of their reciprocals;

\item $X$ computes, over distinct and disconnected endpoint nodes $i,j\in R^\star$ with intermediate node in $N_a\cup\{a\}$, the EBC partial sum. That is: $S_X = \sum_{i,j\in R^\star: A_{ij}=0, j>i} 1 / A_{ij}^2$; and

\item $X$ completes computation of \EBC{a} as $S_X + S_{XY} + S_Y$.

\end{enumerate}
\end{protocol}

\begin{remark}\label{rem:disclosure}
We now briefly comment on where edge privacy is potentially breached, thereby highlighting challenges faced by any solution to Problem~\ref{prob:pebc}.
 When $X$ sends $Y$ its set of neighbours $R^\star$ of $a$, party $Y$ learns directly of all edges incident to $a$ in $X$. 
When $Y$ sends $X$ its 2-path counts $T_{ij}$, while counts aggregate exact connectivity at worst this level of aggregation could be very small therefore revealing information about connections to $a$ within $Y$, and the inter-connections 
between these nodes.
A worst case occurs when there are two nodes in $Y$ connected to $a$: as soon as  $X$ receives the vector of counts of 2-paths spanning $X,Y$, it can learn whether these two nodes are connected.
\end{remark}

\section{A Privacy-Preserving Protocol}  \label{sec:proposed}

We now develop our protocol for private EBC which involves a series of differentially-private mechanisms for 
overcoming the privacy disclosures identified in Remark~\ref{rem:disclosure}. We relegate proofs to the 
\ifdefined\FULLPAPER
Appendices.
\else
full report~\cite{supplemental}.
\fi

We use the exponential mechanism (\viz Lemma~\ref{lem:exponential}) to release a set $R$ of nodes in $X$ that privately approximates $a$'s ego network $R^\star$ in $X$. This is  Protocol~\ref{prot:NonPrivate}.\ref{step:nonPrivateProtForward}'s `forward message'  (Section~\ref{sec:forward}).

While our application of the exponential mechanism protects edge privacy for $X$, there is still potential for privacy disclosure when $Y$ communicates counts to $X$. To overcome this problem, we leverage the Laplace mechanism to privatise vectors of 2-path counts communicated by $Y$ within Protocol~\ref{prot:NonPrivate}.\ref{step:nonPrivateProtBack1}'s `backward message' (Section~\ref{sec:backward}). The components of this message are indexed (in part) by the approximate $R\approx R^\star$ sent in the forward message. A second  Laplace mechanism makes private the partial EBC of Protocol~\ref{prot:NonPrivate}.\ref{step:nonPrivateProtBack2}'s `backward message'.

In this way our privacy-preserving protocol follows the broad-brush sequence of steps outlined in Section~\ref{sec:nonprivate} but is made more involved by the addition of differential privacy.

\subsection{Forward Message}\label{sec:forward}

The goal of the forward message, is to communicate a privacy-preserving approximation to $R^\star = N_a\cap V_X$ chosen from power set $\Pow{V_X\backslash\{a\}}$. In order to leverage the exponential mechanism (\viz Lemma~\ref{lem:exponential}) we must specify a quality function of the form $q: \{0,1\}^{|V_X|^2}\times\Pow{V_X\backslash\{a\}} \to \mathbb{R}$. That is a mapping from the $V_X\times V_X$ adjacency matrix for $E_X$ and a candidate response $R\subseteq V_X\backslash\{a\}$, to a score reflecting the approximation quality of $R^\star$ by $R$. Since the response set is finite (albeit exponential in the graph size), the exponential mechanism then has normalised response probability mass function, 
\begin{IEEEeqnarray}{rCl}
\frac{\exp( q(R) \cdot \epsilon / (2\cdot \Delta q))}{\sum_{S\subseteq X}\exp( q(S) \cdot \epsilon / (2\cdot \Delta q))}\enspace,
\label{eq:exp-response}
\end{IEEEeqnarray}
with implicit dependency on fixed $E_X$ adjacency matrix. 

In designing an appropriate quality function, we typically want $q$ to be maximised uniquely by the desired non-private output $R^\star$. The function should also be a semantically meaningful `distance' between outputs $R$ and $R^\star$ such that the utility bounds for the exponential mechanism of \cite[Lemma 7 and Theorem 8]{mcsherry2007mechanism} make meaningful guarantees. The utility guarantee states that with high probability the released random $R$ has $q$ score not too much lower than $\max_{S\subseteq V_X\backslash\{a\}} q(S)$. And so if $R^\star$ meets this global maximum, then we have that the released set has score not much lower than that of $R^\star$. If responses close in $q$ are also `close' then this guarantees a good approximation to $R^\star$ with high probability.

\begin{remark}
A natural choice for quality function is $q(R)=|R\cap R^\star|$ as it is clearly maximised by $R^\star$. 
However it is not uniquely maximised, indeed for any superset $R^\star\subset S\subseteq V_X\backslash\{a\}$ (including the entire set of nodes) we have that $q(S)=|S\cap R^\star|=|R^\star|$ also. There are many such sets: $2^{|V_X|-1-|R^\star|}$ which is not far from the number of all possible responses $2^{|V_X|-1}$ for modest ego network sizes, in which case the exponential mechanism does not achieve our goal.
\end{remark}

Section~\ref{sec:quality} develops a sound choice of quality function in the symmetric set difference.

In the rest of this section we will abuse notation and abbreviate $X=V_X$ with the meaning understood from context. We will also denote by $\Xa=V_X\backslash\{a\}$.

\subsubsection{Symmetric Set Difference}
\label{sec:quality}

 We adopt (for minimisation) the symmetric set difference with $R^\star$ given by $(R\backslash R^\star)\cup(R^\star\backslash R)$ as a promising basis for quality function design.  Define  complements   $\overline{B}=\Xa\backslash B$ relative to \Xa. We have
\begin{IEEEeqnarray*}{rCl}
(R\backslash R^\star)\cup(R^\star\backslash R) &=& \Xa \backslash \left((R\cap R^\star) \cup \overline{R\cup R^\star}\right)\; ;\; \mbox{and} \\
|(R\backslash R^\star)\cup(R^\star\backslash R)| &=& |\Xa| - |R\cap R^\star| - |\overline{R\cup R^\star}|\enspace,
\end{IEEEeqnarray*}
where the second equality follows from a disjoint union in the first equality's right-hand side. 

In minimising the symmetric difference, dismissing the constant $|\Xa|$ as redundant to optimisation, we can equivalently maximise\footnote{While $q$'s dependence on $R^\star$ is suppressed, it should be implicitly understood.}
\begin{IEEEeqnarray}{rCl}
q(R) &=& |R\cap R^\star| + |\overline{R\cup R^\star}|\enspace. \label{eq:quality}
\end{IEEEeqnarray}
This quality function takes values in $\{0,\ldots,|\Xa|\}$ and is uniquely maximised by $R^\star$. 

While the machinery of the exponential mechanism only requires sensitivity of this quality function (Section~\ref{sec:exp-sensitivity}) to guarantee differential privacy, a significant challenge is involved in sampling from the mechanism's response distribution as it is defined over an enormous response space: the power set of \Xa. Thanks to the amplification of $q$ by the exponential, the distribution's mass varies an incredible amount even for graphs of modest size.

\subsubsection{Equi-Quality Responses}

It will be useful to consider the sets of candidate exponential mechanism responses, with equal quality score value $i$,
\begin{IEEEeqnarray*}{rCl}
Q_i &=& \left\{R \subseteq \Xa : q(r) = i\right\}\enspace,
\end{IEEEeqnarray*}
where $i\in\{0,\ldots,|\Xa|\}$. It can be shown that the $Q_i$ form a partition of $\Pow{\Xa}$: the sets are pairwise disjoint, and their union is all subsets of \Xa. It can also be shown that for $i\in\{0,\ldots,|\Xa|\}$,
$$
|Q_i| = \sum_{k=0}^i 
    \binom{|R^\star|}{k}  \binom{|\Xa|-|R^\star|}{i - k}\nonumber \\
= \binom{|\Xa|}{i}.
$$
A consequence of this identity is an efficient approach to computing the normalising constant for the exponential mechanism response distribution. The proof can be found in the \ifdefined\FULLPAPER Appendices. \else full report~\cite{supplemental}. \fi

\begin{corollary}\label{cor:normaliser}
Consider the normalised exponential mechanism response distribution~\eqref{eq:exp-response} for the quality function given in~\eqref{eq:quality}. The normalising constant is equivalent to  
\begin{IEEEeqnarray*}{rCl}
C &=& \sum_{i=1}^{|X|} |Q_i| \exp\left(\frac{i\cdot\epsilon}{2\Delta}\right) \enspace.
\end{IEEEeqnarray*}
Computing this expression takes time and space $O(|X|)$. 
\end{corollary}

Note that other phases of our protocol, to be specified, also require linear space.
By comparison computing $C$ na\"ively  would take time exponential in $X$ and constant space.

\subsubsection{Two-Stage Sampling}

We will now outline how to sample from the exponential mechanism response distribution, using a two-stage sampling process that delivers a simple approach to implement, and exponential savings in time and space vs na\"ive sampling from the exponential mechanism. 

\begin{remark}
Another standard approach to sampling from challenging distributions is acceptance-rejection sampling or the Metropolis algorithm. However no clear surrogate probability mass presents itself that yields acceptable rates of rejection for practical applicability. 
\end{remark}

\paragraph{Stratified Sampling}
To simplify notation, let us consider the problem at-hand more generally: let $V\in\mathcal{V}$ be a discrete random variable on finite probability space $(p,\mathcal{V})$ \ie a multinomial with $p: v\in\mathcal{V}\mapsto\Pr(V=v)$. Suppose there exists a partition of $\mathcal{V}$ into the disjoint union of $\mathcal{V}_0,\ldots,\mathcal{V}_k$, such that for all $i$, and all $v,v'\in\mathcal{V}_i$, $p(v)=p(v')$ \ie the probability mass  is constant within each part.
We can exactly sample from $V\sim p$ by (1) drawing a random variable that selects a part in the partition according to the relative part sizes then (2) sampling uniformly from within the chosen part---
this is the approach taken by Algorithm~\ref{alg:forward} \ForwardMessage. 
Denote by $p_i$ the constant probability mass $p(v)$ of any $v\in\mathcal{V}_i$. The following result is proven in the  \ifdefined\FULLPAPER Appendices. \else full report~\cite{supplemental}. \fi

\begin{lemma}\label{lem:stratified}
Define random variable $I\in \{0,\ldots,k\}$ where $\Pr(I=i)\propto |\mathcal{V}_i|\cdot p_i$ for each $i\in \{0,\ldots,k\}$, and $U\sim \Unif{\mathcal{V}_I}\mid I$. Then $\Pr(V=v)=\Pr(U=v)$ for all $v\in\mathcal{V}$. Moreover, the probability mass stated for $\Pr(I=i)$ is already normalised, \ie $\Pr(I=i) = |\mathcal{V}_i| \cdot p_i$.
\end{lemma}

\begin{corollary}\label{cor:two-stage}
The following sampling process, implemented in \ForwardMessage Algorithm~\ref{alg:forward}, is equivalent to sampling $R$ from the exponential mechanism~\eqref{eq:exp-response}

    \noindent (i) Sample $I\in\{0,\ldots,|\Xa|\}$ with log-space probability mass
    \begin{align*}
   & \log \Pr(I=i) \\ =&
\begin{cases} 
    -|\Xa|\log\left(1+\exp\left(\frac{\epsilon}{2\Delta}\right)\right), & i = 0\\
    \log(|\Xa|-i+1) - \log i + \log\Pr(I=i-1), & \mbox{o.w.}
\end{cases}
\end{align*}

   \noindent (ii) Sample $R\sim \Unif{Q_I}\mid I$
\end{corollary}

\begin{IEEEproof}
As the exponential mechanism~\eqref{eq:exp-response} on $R\in\Pow{\Xa}$, with quality function~\eqref{eq:quality}, is a multinomial distribution with strata of constant probability given by the $Q_i$, Lemma~\ref{lem:stratified} establishes that the stratified sampler successfully implements the mechanism. All that remains is to compute the probability of 
selecting $Q_i$ as the stratum's cardinality times constant probability (normalised by $C$ as given in Corollary~\ref{cor:normaliser}).
$$    \Pr(I=i) \; = \; |Q_i|\cdot\frac {\exp\left(\frac{i\cdot\epsilon}{2\Delta}\right)}{C} 
      \; = \;   \frac{\tbinom{|\Xa|}{i}  \exp\left(\frac{i\cdot\epsilon}{2\Delta}\right) }
      {\left(1+ \exp\left(\frac{\epsilon}{2\Delta}\right)\right)^{|\Xa|}}\enspace,
      $$
where the final equality follows from the observation that the expression for $C$ is a Binomial expansion.
We simplify and convert this expression to log-space as the expression exponentially increases in $i$: 
\begin{align*}
& \log \Pr(I=i) \\ 
=& \log \binom{|\Xa|}{i} +\frac{i\cdot\epsilon}{2\Delta}- |\Xa|\log\left(1+\exp\left(\frac{\epsilon}{2\Delta}\right)\right)\\
   =&
\begin{cases} 
    -|\Xa|\log\left(1+\exp\left(\frac{\epsilon}{2\Delta}\right)\right), & i = 0\\\
    \log(|\Xa|-i+1) - \log i + \log\Pr(I=i-1), & i > 0,
\end{cases}
\end{align*}
completing the result.
\end{IEEEproof}

\begin{algorithm}[tb]
 \caption{\ForwardMessage Two-Stage Sampler \label{alg:forward}}
 \begin{algorithmic}[1]
 \renewcommand{\algorithmicrequire}{\textbf{Input:}}
 \REQUIRE edge set $E_X$; ego node $a\in V_X$; $\epsilon, \Delta>0$
  \STATE $R^\star \longleftarrow \{v\in V_X \mid \{v,a\}\in E_X\}$
  \STATE $\Xa \longleftarrow V_X\backslash\{a\}$
  \STATE $I \longleftarrow \ITSampler(|\Xa|, \epsilon, \Delta)$
  \STATE $R \longleftarrow \PFSampler(\Xa, R^\star, I)$
  \RETURN{$R$}
  \end{algorithmic}
\end{algorithm}

\subsubsection{Linear-Complexity Sampling}\label{sec:efficient-sampling}

While Corollary~\ref{cor:two-stage} reduces the
problem from sampling from a large support set with
highly skewed probability mass, we must address 
efficient implementation of the two-stage sampling.

\begin{algorithm}[tb]
 \caption{\ITSampler \label{alg:itsampler}}
 \begin{algorithmic}[1]
 \renewcommand{\algorithmicrequire}{\textbf{Input:}}
 \REQUIRE cardinality $|\Xa|$; $\epsilon, \Delta>0$ \\
 \texttt{// Compute log-space PDF of $I$}
  \STATE $p_0 \longleftarrow -|\Xa|\log\left(1 + \exp\left(\frac{\epsilon}{2\Delta}\right)\right)$
  \FOR{$i\in [|\Xa|]$}
    \STATE $p_i \longleftarrow p_{i-1} + \log(|\Xa|-i+1) - \log i$
  \ENDFOR \\
  \texttt{// Search in CDF for random quantile}
  \STATE $\psi \longleftarrow - \mathrm{Exp}(1)$
  \STATE $c \longleftarrow p_0$
  \FOR{$I\in [|\Xa|]$}
    \IF{ $c\geq \psi$ }
      \RETURN{$I - 1$}
    \ENDIF
    \STATE $c \longleftarrow \log(\exp(c) + \exp(p_I))$
  \ENDFOR
  \RETURN{$I$}
\end{algorithmic}
\end{algorithm}

\begin{algorithm}[tb]
 \caption{\PFSampler \label{alg:pfsampler}}
 \begin{algorithmic}[1]
 \renewcommand{\algorithmicrequire}{\textbf{Input:}}
 \REQUIRE node set $\Xa$; node set $R^\star\subseteq\Xa$; $I\in\{0,\ldots, |\Xa|\}$ 
 \STATE $R \longleftarrow R^\star$
 \STATE $V_1,\ldots, V_{|\Xa|-I} \sim \Unif{\Xa}$ without replacement
 \FOR{$j\in [|\Xa| - I]$}
   \IF{$V_j\in R$} 
     \STATE $R \longleftarrow R\backslash\{V_j\}$
   \ELSE
     \STATE $R \longleftarrow R\cup\{V_j\}$
   \ENDIF
 \ENDFOR
 \RETURN{$R$}
\end{algorithmic}
\end{algorithm}

\paragraph{Inverse Transform Sampling of $I$} 
The sampling of multinomial $I$ over much smaller
support set $\{0,\ldots,|\Xa|\}$ can be accomplished efficiently via \term{inverse transform sampling}. Given access to a random variable $Z$'s (invertible) CDF $F_Z$, one can sample realisations of $Z$ by first sampling $\psi\sim\Unif{[0,1]}$ then releasing  quantile $F_Z^{-1}(\psi)$. For $Z\in\mathbb{Z}$, we can always take
$F_Z^{-1}(\psi) = \inf\{z\in\mathbb{Z} \mid F_z(z)\geq\psi\}$.
However highly-skewed distributions can suffer from numeric instability in floating-point computation of the CDF. Though not as severe as for $R$, this remains a problem for $I$. To combat this we employ a library for arbitrary floating-point precision in our implementation (see Section~\ref{sec:experiments}) and we represent $I$'s probability mass in log space as reported in Corollary~\ref{cor:two-stage}. In this case, the inverse transform sampler is easily adapted as proved in the \ifdefined\FULLPAPER Appendices. \else full report~\cite{supplemental}.\fi

\begin{prop}\label{prop:sampling-I}
Consider r.v. $I$ defined in Corollary~\ref{cor:two-stage}. It can be sampled via the \ITSampler (Algorithm~\ref{alg:itsampler})
in time and space $O(|\Xa|)$ 
given
oracle access to exponential r.v.'s.
\end{prop}

\paragraph{Pick-and-Flip Sampling of $R$} 
After sampling $I$, we must 
sample
$R$ uniformly from within 
chosen stratum $Q_I$, a constrained and potentially large subset of $\Pow{\Xa}$. 
For sampled $R$ we are to have $q(R) = |R\cap R^\star| + |\overline{R\cup R^\star}| = I$, so the size of $R$ and $R^\star$'s  symmetric set difference is 
$|\Xa| - I$. Moreover, this describes all candidates for $R$ within $Q_I$.

\begin{prop}\label{prop:sampling-U}
If r.v. $I$ is sampled with \ITSampler (Algorithm~\ref{alg:itsampler}), then Algorithm~\ref{alg:pfsampler} \PFSampler 
yields a sample of r.v. $R$ defined in Corollary~\ref{cor:two-stage} in time and space
$O(|\Xa|)$.
\end{prop}

\begin{IEEEproof}
Every time a new node $V_j$ sampled from $\Xa$, it could be sampled from either $\Xa\backslash R^\star$ or $R^\star$. In both cases, $V_j$ is added to the set difference. This loop invariance continues until the set difference size reaches $|\Xa|-I$ establishing $R\in Q_I$ and uniformly so due to the uniform sampling of the $V_j$. 
Sampling the nodes (like the rest of the algorithm) can be achieved in linear time/space with Fisher-Yates shuffling.
\end{IEEEproof}

\subsubsection{Quality Function Global Sensitivity}\label{sec:exp-sensitivity}

The remaining ingredient for invoking the exponential mechanism to privately release $R\approx R^\star = N_a \cap V_X$, is bounding $q$'s sensitivity.

\begin{lemma}\label{lem:exp-sensitivity}
Consider any fixed $R\subseteq\Xa$ and $X$-contained ego networks $R^\star$, ${R^\star}'$ induced by neighbouring adjacency matrices on $E_X$ and fixed ego node $a$. Noting explicitly the dependence of the quality function \eqref{eq:quality} on non-private ego network, $|q(R,R^\star) - q(R,{R^\star}')|\leq 1$.
\end{lemma}

\begin{IEEEproof}
Consider the effect of switching an edge within $X$ on the symmetric difference cardinality between $R, R^\star$ \ie the
quality function. 
Adding/removing an edge can impact at most one node being neighbours with ego node $a$; it can therefore only decrease or increase the first or second terms of $q$ by 1, at most. Since these two sets are disjoint, it cannot change both simultaneously.
\end{IEEEproof}

\begin{theorem}\label{thm:forward}
\ForwardMessage (Algorithm~\ref{alg:forward})
takes time and space $O(|\Xa|)$, and when run
with $\Delta=1$, preserves
$\epsilon$-differential privacy of the
edge set $E_X$ within party $X$'s network.
\end{theorem}

\begin{IEEEproof}
Privacy follows from Lemma~\ref{lem:exponential},  Corollary~\ref{cor:two-stage} and Lemma~\ref{lem:exp-sensitivity},
complexity from Propositions~\ref{prop:sampling-I} and~\ref{prop:sampling-U}.
\end{IEEEproof}

\subsection{Backward Message}\label{sec:backward}
Analogous to the non-private protocol of Section~\ref{sec:nonprivate} (Figure~\ref{fig:EBC}), $Y$ receives node-set $R$ approximating $a$'s ego network in $X$, via \ForwardMessage. Subsequently, $Y$ must send back: counts $T_{ij}$ of $2$-paths spanning $X,Y$ with intermediate node in $Y$---indexed by $R$; and its partial EBC sum $S_Y$ over paths with endpoints in $Y$ and intermediate node in $R\cup\{a\}\cup (N_a\cap V_Y)$. Note this last set is the private appoximation to $N_a\cup\{a\}$. We apply in \BackwardMessage (Algorithm~\ref{alg:backward}) the Laplace mechanism (Lemma~\ref{lem:laplace}) to both backward message components to avoid disclosure of edges $E_Y$ in $Y$.

Note cases in which \BackwardMessage need not be run by $X$:
If $| N_a\cap V_Y| \leq 1$ there are no paths incident to $Y$ with intermediate point in $Y$; or contained entirely within $Y$. We may therefore assume within the algorithm that these cases are not present.

\begin{algorithm}[tb]
 \caption{\BackwardMessage}\label{alg:backward}
 \begin{algorithmic}[1]
 \renewcommand{\algorithmicrequire}{\textbf{Input:}}
 \REQUIRE 
   ego node $a\in V_X$; edge sets $E_{XY}, E_Y$; private node set $R\subseteq V_X\backslash\{a\}$; $\epsilon, \Delta_1, \Delta_2>0$ \\
  \texttt{// Count 2-paths of type $X-Y-Y$}
  \FOR{$i\in R$ and $j\in N_a\cap V_Y $}
  \STATE $K \longleftarrow \{k\in N_a\cap V_Y \mid \{i,k\}\in E_{XY}, \{k,j\}\in E_Y\}$
 \STATE $T_{ij} \longleftarrow |K| + \mathrm{Lap}(2 \Delta_1/\epsilon)$
 \ENDFOR \\
 \texttt{// Partial EBC sum over 2-paths in $Y$}
 \STATE $E_Y'\longleftarrow E_{XY}\cup E_Y$
 \STATE $S_Y\longleftarrow0$.
 \FOR{$i,j\in N_a\cap V_Y$ with $j>i$ and $\{i,j\}\notin E_Y$}
   \STATE $K\longleftarrow\{k\in R\cup\{a\}\cup (N_a\cap V_Y) \mid \{i,k\}, \{k,j\}\in E_Y'\}$
   \STATE $S_Y \longleftarrow S_Y + \frac{1}{|K|}$
 \ENDFOR
 \STATE $S_Y \longleftarrow S_Y + \mathrm{Lap}(2 \Delta_2 / \epsilon)$
 \RETURN{$\mathbf{T}, S_Y$}
\end{algorithmic}
\end{algorithm}

\begin{algorithm}[tb]
 \caption{\PrivateEBC}\label{alg:privateEBC}
 \begin{algorithmic}[1]
 \renewcommand{\algorithmicrequire}{\textbf{Input:}}
 \REQUIRE node sets $V_X$ and $V_Y$; $T$; $S_Y$; ego node $a\in V_X$; edge sets $E_{XY}, E_Y$\\ 
\texttt{// Party $X$ runs:} 
 \STATE $R \longleftarrow$ {\footnotesize\ForwardMessage}$(E_X, a, \epsilon, \Delta_0)$ \\
 \STATE Send $R$ to party $Y$ \\
 \texttt{// Party $Y$ runs:}
 \STATE $\mathbf{T}, S_Y\longleftarrow$~{\footnotesize\BackwardMessage}$(a,E_{XY},E_Y,R,\epsilon,\Delta_1,\Delta_2)$
 \STATE Send $\mathbf{T}, S_Y$ to party $X$ \\
 \texttt{// Party $X$ runs:}
 \STATE $S_{XY} \longleftarrow 0$
 \FOR{$i\in R^\star$ and $j\in N_a\cap V_Y$ where $\{i,j\}\notin E_{XY}$}
   \STATE $K \longleftarrow \{k\in R^\star\cup\{a\} \mid \{i,k\}\in E_X, \{k,j\}\in E_{XY} \}$
   \IF{$i\notin R$} 
     \STATE $T_{ij} \longleftarrow 0$
   \ENDIF
   \STATE $T_{ij} \longleftarrow T_{ij} + |K|$
   \STATE $S_{XY} \longleftarrow S_{XY} + \frac{1}{T_{ij}}$
 \ENDFOR
 \STATE $S_X \longleftarrow 0$
 \FOR{$i, j\in R^\star$ where $j>i$ and $\{i,j\}\notin E_X$}
   \STATE $K \longleftarrow \{k\in N_a \cup\{a\} \mid \{i,k\}\in E_X, \{k,j\}\in E_X \}$
   \STATE $S_X \longleftarrow S_X + \frac{1}{|K|}$
 \ENDFOR
 \RETURN $S_X+S_{XY}+S_Y$ to party $X$
 \end{algorithmic}
\end{algorithm}

\subsubsection{Privately Counting Paths}
The first part of the backward message compares the $T_{ij}$---noisy counts of 2-paths with intermediate node in $N_a\cap V_Y$---over $i$ in the given $R$ approximating $R^\star$, and $j\in N_a\cap V_Y$. 
The sensitivity of these counts relates to adding or removing an edge in $E_{XY}$, as
follows, with proof given in the
\ifdefined\FULLPAPER Appendices. \else full report~\cite{supplemental}. \fi

\begin{lemma}
\label{lem:fsensitivity}
Let query $f$ denote the vector-valued non-private response $\mathbf{T}$. 
The $L_1$-global sensitivity of $f$ is upper-bounded by $\Delta f=2|R|$.
\end{lemma}

\captionsetup[subfigure]{subrefformat=simple,labelformat=simple,listofformat=subsimple}
\renewcommand\thesubfigure{(\alph{subfigure})}
\begin{figure*}[t]
\centering
\subfloat[Average relative error of the 60 random nodes with $\epsilon=0.1$ to  $7$, Facebook data set.]{\label{fig:refa}\includegraphics[width=0.32\textwidth, keepaspectratio]{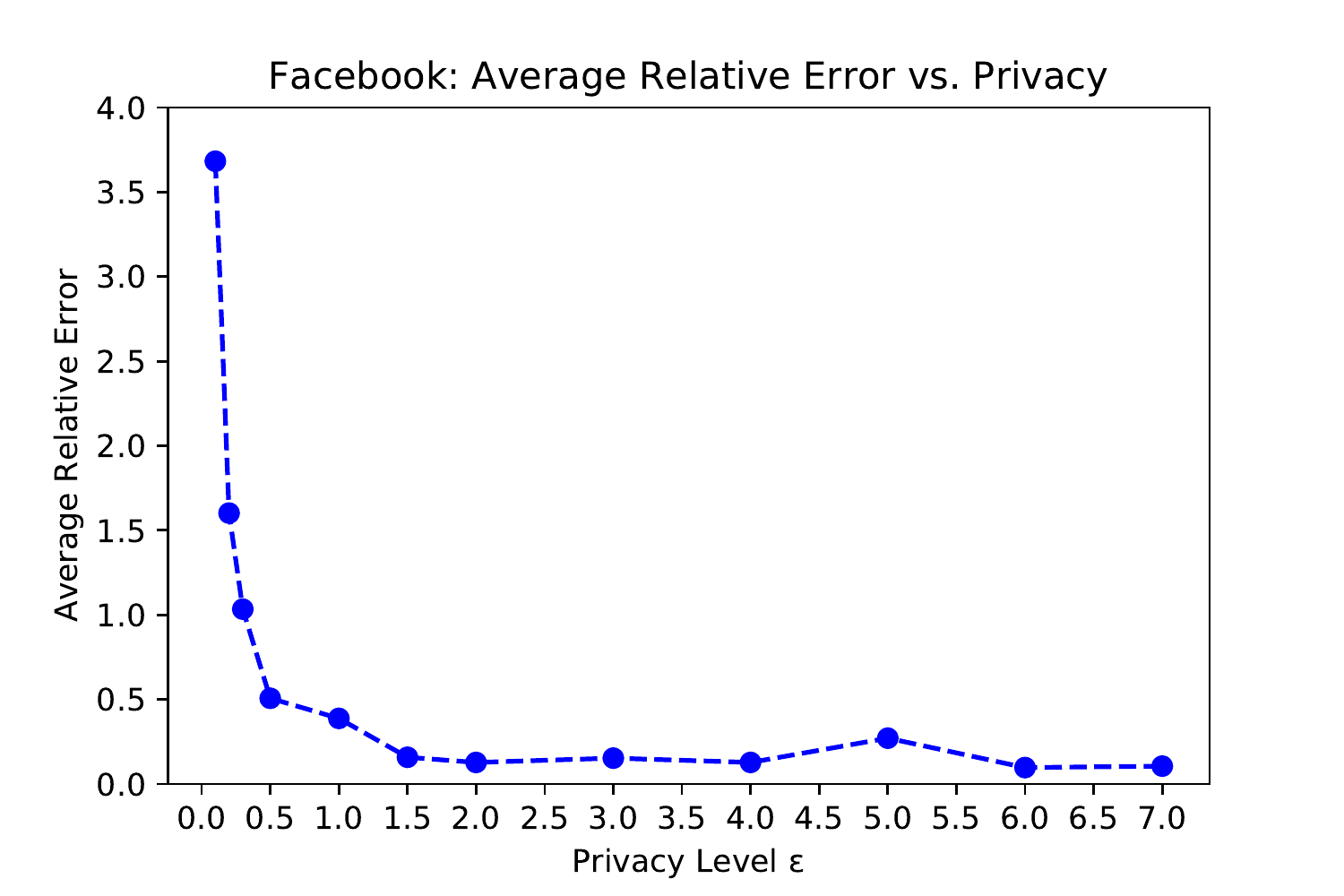}
}
\hfill
\subfloat[Average relative error of 60 nodes with $\epsilon=0.1$ to $7$, Enron data set.]{\label{refb}\includegraphics[width=0.32\textwidth, keepaspectratio]{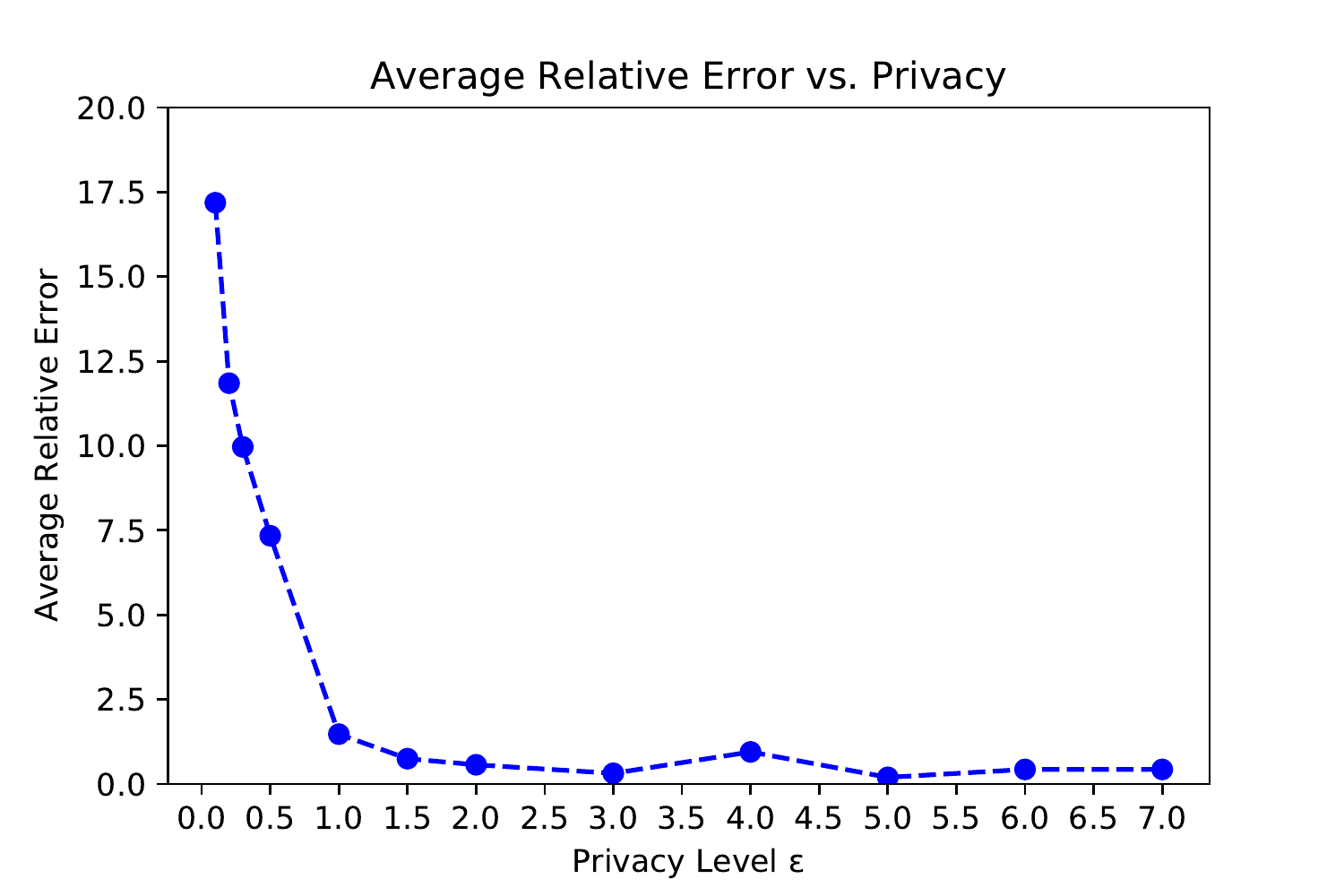}
}
\hfill
\subfloat[Average relative error of 60 nodes with $\epsilon=0.1$ to $7$, PGP data set]{\label{refc}\includegraphics[width=0.32\textwidth,keepaspectratio]{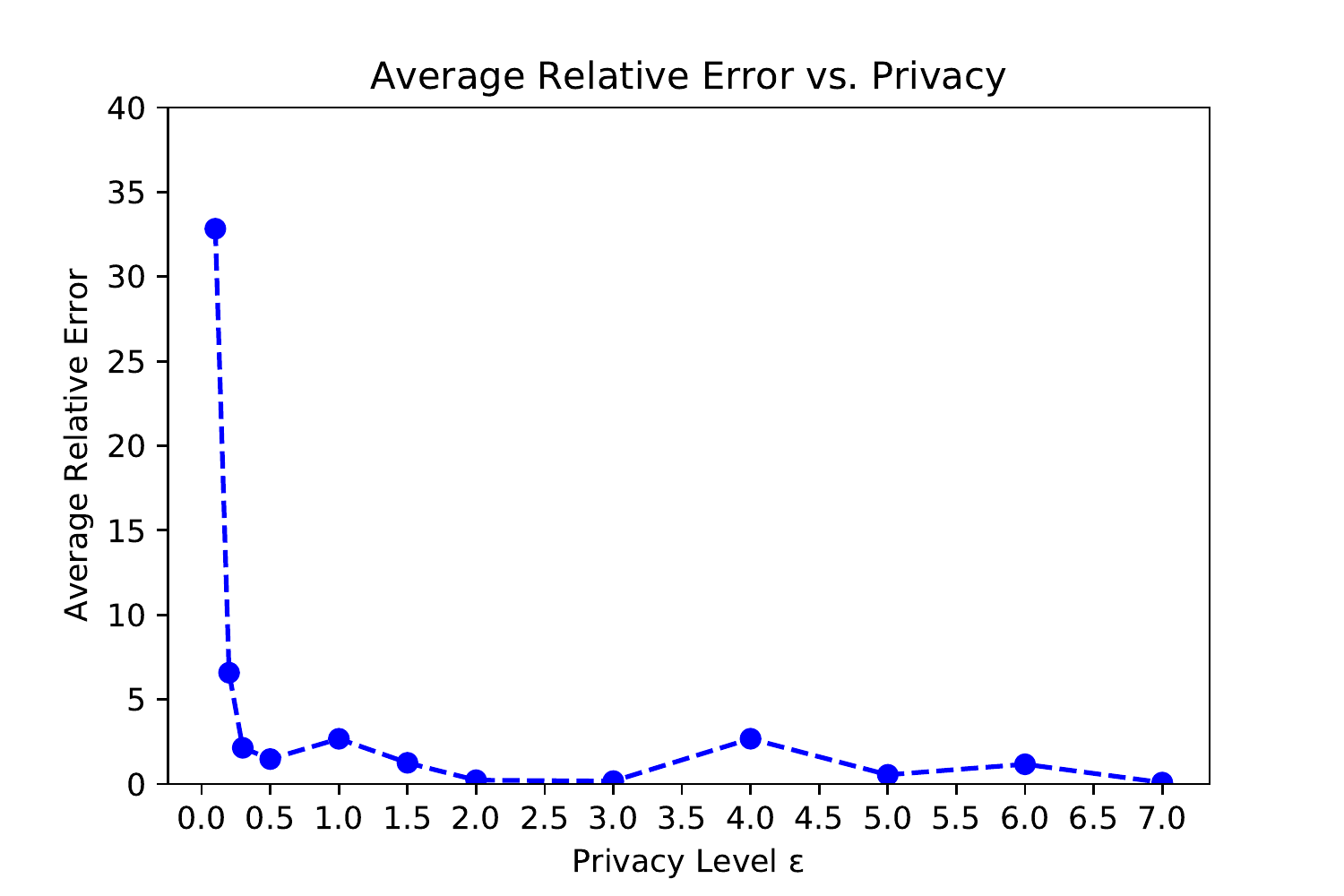}}

\subfloat[Average relative error of 60 nodes in three different mechanisms, $\epsilon=1$, Facebook data set.]{\label{refd}\includegraphics[width=0.32\textwidth, keepaspectratio]{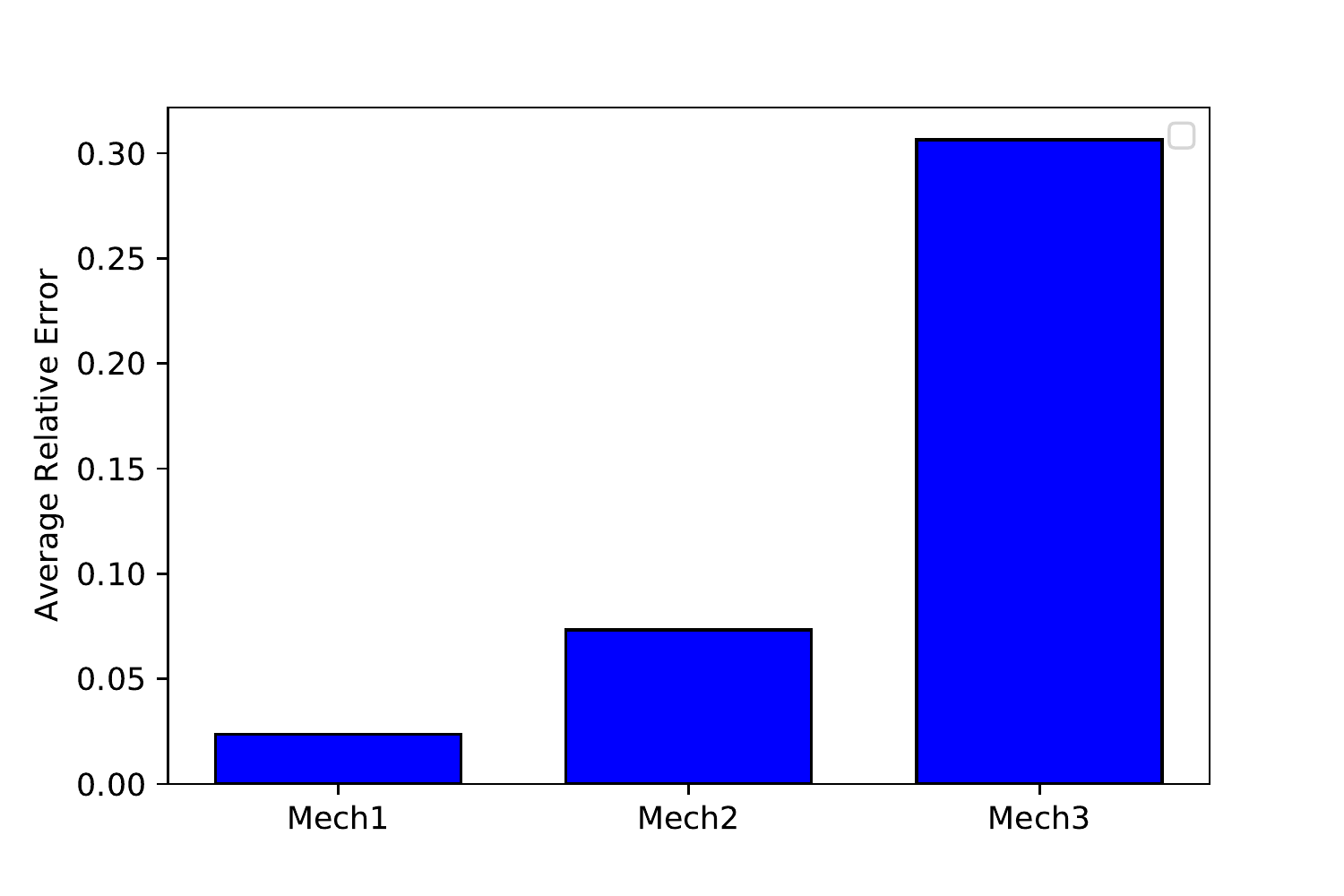}}
\hfill
\subfloat[Time of computing 20 random nodes with $\epsilon=0.1$ to $7$, Facebook data set. ]{\label{refe}\includegraphics[width=0.32\textwidth, keepaspectratio]{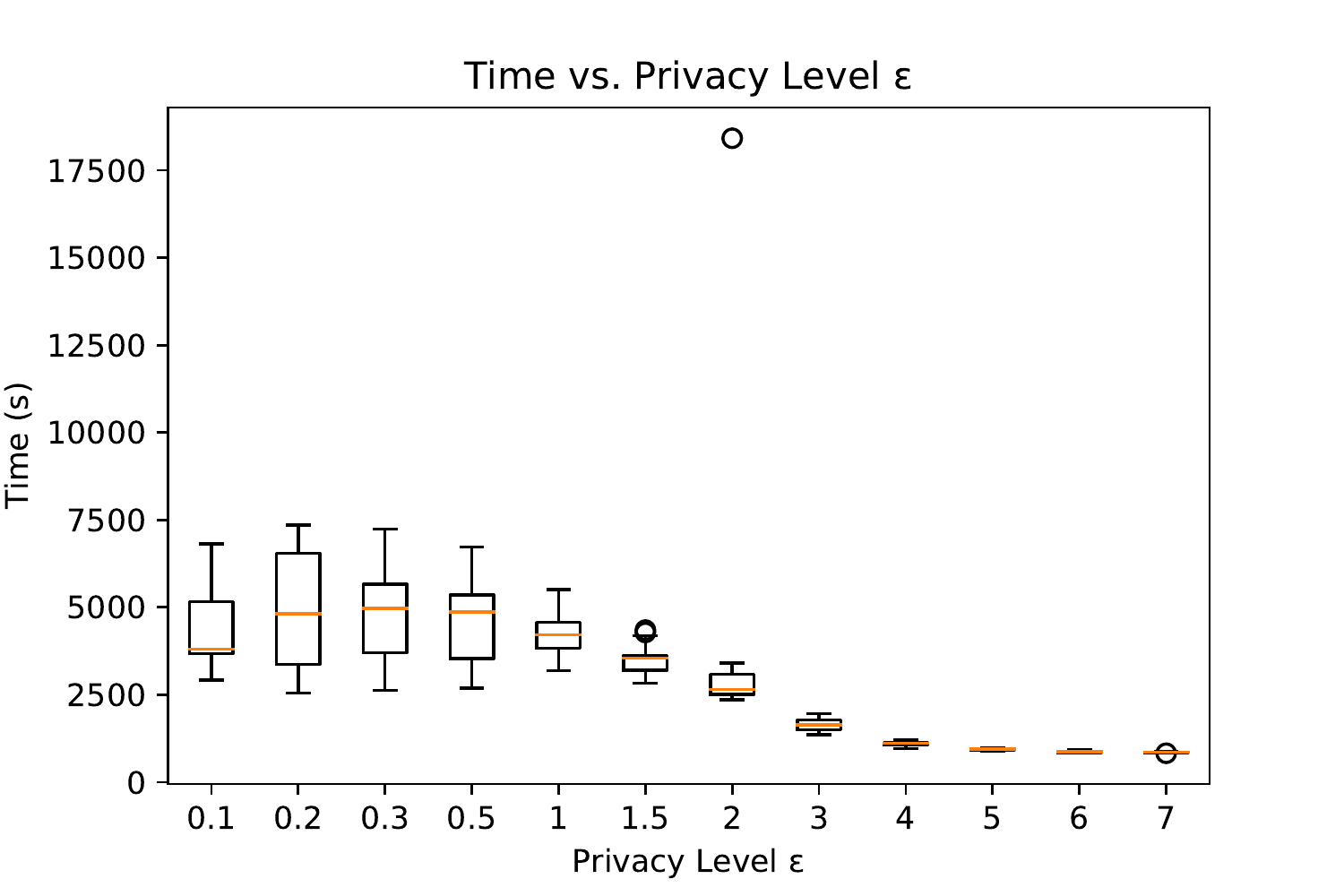}
}
\hfill
\subfloat[Relative error of 100 nodes with different degrees for $\epsilon=1$, Facebook data set.]{\label{reff}\includegraphics[width=0.32\textwidth, keepaspectratio]{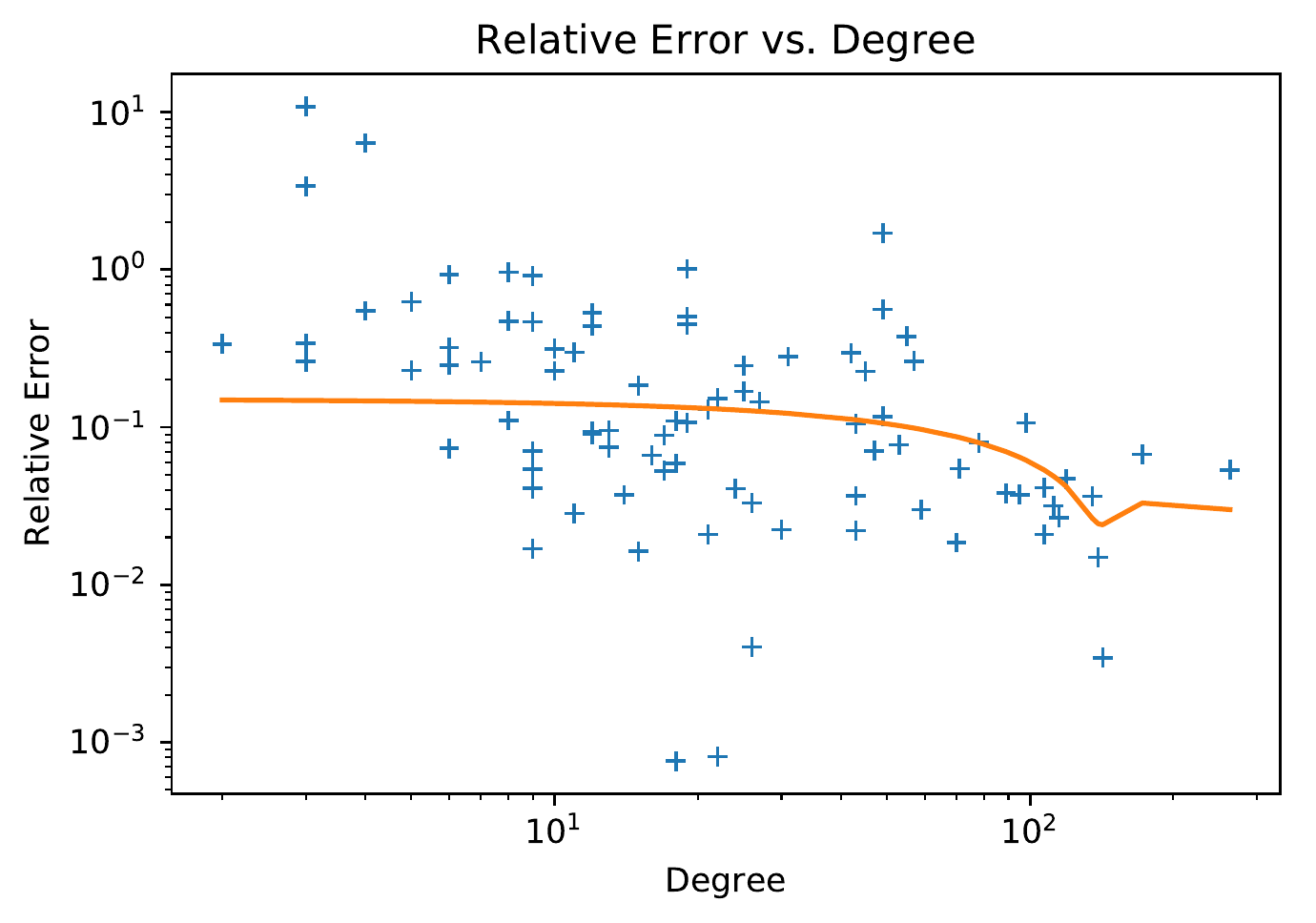}
}
\caption{Experimental results for Facebook, Enron and PGP data sets.}
\label{fig_sim}
\end{figure*}



\subsubsection{Private Partial EBC}

The second part of \BackwardMessage is a partial EBC sum over 2-paths with end-points in $N_a\cap V_Y$ (and intermediate point in either the same or $R\cup\{a\}$) as in  Protocol~\ref{prot:NonPrivate}.\ref{step:nonPrivateProtBack2}.  
We again apply the Laplace mechanism to avoid privacy disclosure of edges in $Y$, which requires bounding sensitivity of the non-private sum as follows. The proof can be found in the \ifdefined\FULLPAPER Appendices. \else full report~\cite{supplemental}. \fi

\begin{lemma}  \label{lem:f-primeSensitivity}
Let query $f'$ denote the partial EBC sum over 2-paths with end-points $i,j\in N_a\cap V_Y$ and intermediate node $k\in (N_a\cap V_Y)\cup R\cup\{a\}$.
Then the $L_1$-global sensitivity of $f'$ is upper-bounded by $\Delta f'= |N_a\cap V_Y|-1$.
\end{lemma}

As both applications of the Laplace mechanism run with privacy budget $\epsilon/2$, Lemma~\ref{lem:composition} implies overall $E_Y$ edge privacy is guaranteed.

\begin{corollary}
\label{cor:backward-privacy}
\BackwardMessage (Algorithm~\ref{alg:backward}) takes time and space $O(\max\{|R|,N\}\cdot N)$ where $N=|N_a\cap V_Y|$, and when run with $\Delta_1=2|R|, \Delta_2=N-1$, preserves $\epsilon$-differential privacy of edge set $E_Y$ within party $Y$'s network.
\end{corollary}

\subsection{\PrivateEBC: Putting it All Together}

After parties $X$ and $Y$ have respectively run \ForwardMessage and \BackwardMessage, $X$ must complete the computation of the private EBC. As shown in Algorithm~\ref{alg:privateEBC}, \PrivateEBC comprises two phases that closely mirror the two components of \BackwardMessage: counting 2-paths spanning $X,Y$, and counting 2-paths with endpoints in $X$. 

Within the first stage we incorporate \BackwardMessage noisy counts $T_{ij}$ contributed by $Y$, which count paths having intermediate nodes in $Y$. Party $X$ simply increments these values with counts of paths having endpoints in $X$. The sum reciprocals forms $S_{XY}$. \emph{We make one optimisation to utility at no cost to privacy: counts $T_{ij}$ for $i\in R\backslash R^\star$ are discarded.}

A straightforward sum of paths with endpoints in $X$ and intermediate points in $N_a\cup\{a\}$ completes $S_X$. Finally party $X$ completes \PrivateEBC by summing the partial EBCs.

\begin{remark}
We opt to use the same $\epsilon$ privacy budget for both parties $X$ (Theorem~\ref{thm:forward}) and $Y$ (Corollary~\ref{cor:backward-privacy}) in \PrivateEBC out of symmetry. However the algorithm can operate with separate budgets if desired.
\end{remark}

\section{Experiments}  \label{sec:experiments}

To empirically validate the effectiveness of \PrivateEBC we ran experiments on three graph data sets: 
a Facebook friendship graph~\cite{konect} with 63,731 vertices and 817,035 edges; the Enron email network~\cite{SNAP} with 36,692 nodes and 183,831 edges; and the Pretty Good Privacy (PGP)~\cite{konect} data set with 10,680 users as vertices and 24,316 inter-user interactions as edges. We follow a random process to partition the nodes, while the structure of the graph stays intact: nodes are assigned to parties $X$ or $Y$ independently and uniformly at random, while edges are not changed.  

The experiments were run on a server with $2\times28$ core Xeon's (112 threads with hyper threading) and 1.5 TB RAM, using Python~3.7 without parallel computations for fair comparison.
We use relative error between true and private EBC---the lower the relative error the higher the utility. We employed the \texttt{Mpmath} arbitrary precision library  and set the precision to 300 bits. Arbitrary precision is vital for implementing inverse transform sampling as described in Section~\ref{sec:efficient-sampling}.

\section{Results}
We first examine the relationship between utility and privacy for \PrivateEBC. For 60 uniformly-at-random ego nodes selected  from randomly-partitioned party $X$, we report average relative error (comparing private and true EBC) for a range of privacy levels $\epsilon$ between 0.1 and 7. 
Figure~\subref*{fig:refa}---\subref*{refc} show the results for the three datasets, where it is apparent that average relative error decreases dramatically when $\epsilon$ is increased to 1, and stays very small for larger $\epsilon$. For $\epsilon> 0.5$, average relative error is usually below 50\%. And at the strong guarantee of $\epsilon=1.5$ the average relative error is 16\% (Facebook) 47\% (Enron) and 25\% (PGP).

As we have employed three different privacy-preserving mechanisms in our proposed protocol---one exponential (Mech1) and two Laplace (Mech2, Mech3)---we examine each separately to evaluate how they affect overall relative error. Specifically, we run the \PrivateEBC protocol with only one of the privacy-preserving mechanisms intact and use the non-private version for remaining mechanisms, with each of Mech1--Mech3 taking turns being private. In this way we can isolate the incremental cost to utility of each mechanism. Figure~\subref*{refd} reports the results on Facebook, which demonstrate that Mech1 \ForwardMessage has the least impact on the relative error while Mech3 \BackwardMessage second component, has the highest impact. This suggest future work may focus on the third mechanism. 

We next report on timing analysis for \PrivateEBC as function of privacy level. Median computation time of 20 random ego nodes for $\epsilon$ from 0.1 to 7 is reported in Figure~\subref*{refe} on Facebook data. Here total time is overall decreasing as privacy decreases (increasing $\epsilon$), while a small increase to runtime can be seen at very high levels of privacy (low but increasing $\epsilon)$. This dual effect is slightly more pronounced on Enron and PGP (see the \ifdefined\FULLPAPER Appendices\else full report~\cite{supplemental}\fi), and is likely due to different behaviours in the protocol with increasing $\epsilon$. When the set difference of $R$ and $R\star$ is small, the two-stage sampler generates just small numbers of nodes in faster time.
However faster runtime with lower privacy dominates behaviour overall. Moreover any effect of privacy is not strong, with at most a $5\times$ change in runtime which across data sets is practical at under 10 min (median) on the larger data sets. 

Figure~\subref*{reff} shows how the relative error between true  and private EBC varies by ego node degree. We report results on $\epsilon=1$, which do not show significant dependence: for node degrees up to $10^2$, deviation is approximately 7\% of the maximum relative error which is low. 

\section{Conclusion and Future Work}

In this paper we have developed the \PrivateEBC algorithm which comprises a protocol of differentially-private mechanisms for cooperative 2-party computation of egocentric betweenness centrality. Theoretical and empirical results demonstrate that our approach achieves strong privacy guarantees for both parties which achieving practical levels of utility with efficient time and space complexity. Notably we contribute a novel two-stage sampler that improves upon the exponential mechanism's time and space complexities exponentially. 
\PrivateEBC should extend naturally to multiple networks---we expect to add to our empirical investigations of efficiency in that case. 
It would be interesting to extend differential privacy to the case in which the answer needs to be returned \emph{by} the party whose node is being queried to some untrusted authority.  

\section*{Acknowledgment}
This work is supported by the Australian Research Training Program and the Australian Research Council DE160100584.
\bibliographystyle{IEEEtran}
	\bibliography{IEEEabrv,references.bib}

\ifdefined\FULLPAPER

\appendix
\section{Supplemental Material}

\subsection{Proof of Corollary~\ref{cor:normaliser}}

Consider the denominator of the exponential response distribution~\eqref{eq:exp-response}:
\begin{eqnarray*}
C &=& \sum_{R\subseteq X}  \exp\left(\frac{q(R)\cdot\epsilon}{2\Delta}\right)             \;=\; \sum_{i=1}^{|X|} \sum_{R\in Q_i} \exp\left(\frac{q(R)\cdot\epsilon}{2\Delta}\right) \\
&=& \sum_{i=1}^{|X|} \sum_{R\in Q_i} \exp\left(\frac{i\cdot\epsilon}{2\Delta}\right)
\;=\; \sum_{i=1}^{|X|} |Q_i| \exp\left(\frac{i\cdot\epsilon}{2\Delta}\right)\enspace,
\end{eqnarray*}
establishing the result.

\subsection{Proof of Lemma~\ref{lem:stratified}}

The proof follows by splitting on $I$, the chain rule of probability, and by definition of the r.v.'s. For any $v\in\mathcal{V}$, denote $i(v)\in \{0,\ldots,k\}$ to be such that  $v\in\mathcal{V}_{i(v)}$, then
\begin{eqnarray}
\Pr(U=v) &=& \sum_{i=0}^k \Pr(U=v, I=i)\nonumber \\
&=& \sum_{i=0}^k \Pr(U=v\mid I=i) \cdot \Pr(I=i)\nonumber \\
&=& \Pr(U=v\mid I=i(v)) \cdot \Pr(I=i(v))\nonumber \\ 
&=& \left(\frac{1}{|\mathcal{V}_{i(v)}|}\right) \cdot \left(\frac{|\mathcal{V}_{i(v)}|\cdot p_{i(v)}}{\sum_{j=0}^k |\mathcal{V}_j|\cdot p_j}\right)\nonumber \\
&=& \frac{p_{i(v)}}{\sum_{j=0}^k |\mathcal{V}_j|\cdot p_j}\nonumber \\
&=& p_{i(v)}\nonumber \\
&=& \Pr(V = v)\enspace\nonumber.
\end{eqnarray}
The penultimate equality follows from
$$
1 = \sum_{v'\in\mathcal{V}} \Pr(V=v') 
= \sum_{i=0}^k \sum_{v'\in\mathcal{V}_i} p_i 
= \sum_{i=0}^k |\mathcal{V}_i| \cdot p_i \enspace.$$
This also establishes that the probability mass $\Pr(I = i)\propto |\mathcal{V}_i|\cdot p_i$ is already normalised.

\subsection{Proof of Proposition~\ref{prop:sampling-I}}

The pseudo-inverse of the CDF follows the general case, while it is easy to show that $-\log Z$, for $Z\sim\Unif{[0,1]}$, is distributed as $\mathrm{Exp}(1)$.  The time complexity corresponds to both computing the CDF (which need not be stored in its entirety) and a linear search for its inversion. 

\subsection{Proof of Lemma~\ref{lem:fsensitivity}}

Suppose  that graphs $G_1$ and $G_2$ differ in some edge $\{j,k\}$  with $j$ and $k$ both in $N_a\cap V_Y$ (that is, the edge $\{j,k\}$ would belong to $E_Y$ ). Our task is to upper bound the corresponding change to counts $\|f(G)-f(G')\|_1$ resulting from running query $f$ on the two graphs. There can be at most $R$ choices of endpoint node $i\in R$ for forming 2-hop paths $(i,k,j)$ affected by the addition/deletion. Similarly there can be at most $R$ choices of endpoint $i\in R$ for paths $(i, j, k)$ affected by the addition/deletion. For each of these $2R$ paths the addition/deletion can affect $\|f(G)-f(G')\|_1$ by at most 1. This proves the result.

\subsection{Proof of Lemma~\ref{lem:f-primeSensitivity}}

 There are two ways the addition or removal of an edge can affect $S_Y$. 
If the edge is the one between endpoints $i$ and $j$, then this can change the $1/A^2_n(i,j)$ term by at most 1, (from 0 to 1, in the case that the only other connection between $i$ and $j$
 is via $a$).  If the edge is within $N_a\cap V_Y$, 
 then it can affect a $1/A^2_n(i,j)$ term by at most $1/2$: the denominator is incremented/decremented by 1 while the denominator must always be at least 1 as a 2-path must go through $a$.  This can occur for at most $2\cdot(|N_a\cap V_Y| - 2)$ terms in the sum, because they're paths involving some intermediate node in $Y$ that is neither $i$ nor $j$.  So overall the change in $S_Y$ resulting from the addition or removal of one edge is at most:
 \begin{eqnarray*}
    \Delta f' &=& 1 + \frac{2\cdot\left(|N_a\cap V_Y| -2\right)}{2} = |N_a\cap V_Y|-1.      
 \end{eqnarray*}

\begin{figure}
\centering
\subfloat[]{\label{refap1}\includegraphics[width=0.32\textwidth, keepaspectratio]{EpsilonVStimeenron.pdf}
}
\hfil
\subfloat[]{\label{refap2}\includegraphics[width=0.32\textwidth, keepaspectratio]{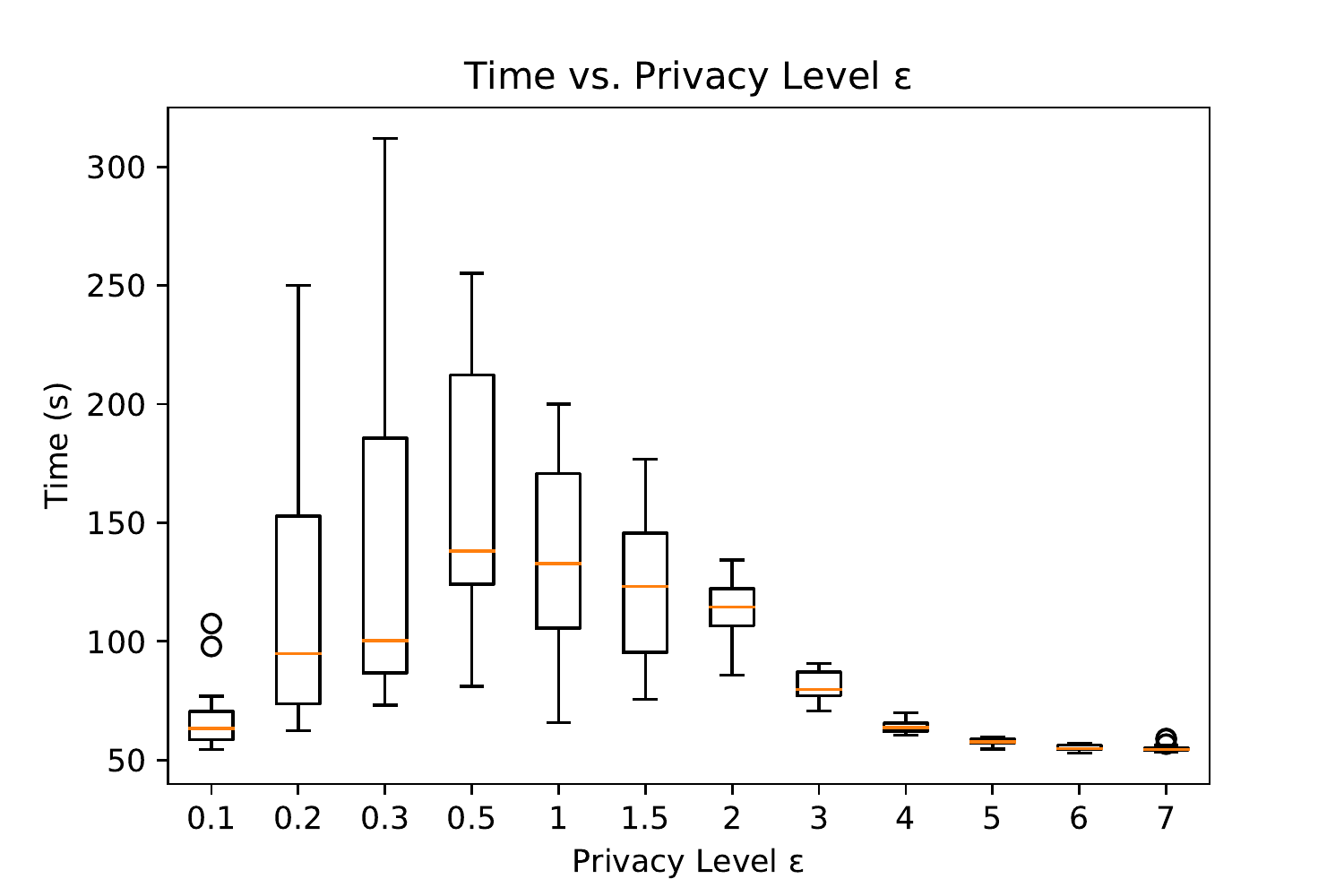}
}
\caption{\protect\subref{refap1}: Time of computing 20 random nodes vs. $\epsilon$ on Enron data set. \protect\subref{refap2}: Time of computing 20 random nodes vs. $\epsilon$ on PGP data set.  }
\label{appendix}
\end{figure}
\subsection{Additional Experimental Results}
We present the timing analyses for \PrivateEBC as function of privacy level. Figure~\subref*{refap1} and~\subref*{refap2} show the results for Enron and PGP data sets which present similar behaviour as the Facebook data \cf Fig.~\ref{refe}.

\fi

\end{document}